\documentclass[12pt]{article}

\usepackage{scicite}
\usepackage{times}
\usepackage{amssymb}
\usepackage{graphicx}
\usepackage{dcolumn}
\usepackage{bm}
\usepackage{amsmath}
\usepackage{mathrsfs}
\usepackage{textcomp}
\usepackage{siunitx}

\usepackage{color}

\usepackage[unicode=true,bookmarks=true,bookmarksnumbered=false,bookmarksopen=false,breaklinks=false,pdfborder={0 0 1},backref=false,colorlinks=true]{hyperref}

\hypersetup{linkcolor=magenta,urlcolor=blue,citecolor=blue,pdfstartview={FitH},hyperfootnotes=false,unicode=true}

\def\be{\begin{equation}}
\def\ee{\end{equation}}
\def\bea{\begin{eqnarray}}
\def\eea{\end{eqnarray}}
\def\nn{\nonumber}

\newenvironment{sciabstract}{%
\begin{quote} \bf}
{\end{quote}}

\title{Dimension Crossing Turbulent Cascade in an Excited Lattice Bose Gas}

\author
{Tianwei Zhou$^{1,5}$, Ruixiao Yao$^{1}$, Kaixiang Yang$^{1}$, Shengjie Jin$^{1}$, \\ 
Yueyang Zhai$^{2}$, Xuguang Yue$^{3}$, Shifeng Yang$^{1}$, Xiaoji Zhou$^{1}$, \\
Xuzong Chen$^{1,\star}$, Xiaopeng Li$^{4,\ast}$
\\
\normalsize{$^{1}$School of Electronics Engineering and Computer Science,}\\
\normalsize{Peking University, Beijing 100871, China}\\
\normalsize{$^{2}$Innovative Research Institute of Frontier Science and Technology,}\\
\normalsize{Beihang University, Beijing 100191, China}\\
\normalsize{$^{3}$State Key Laboratory of Magnetic Resonance and Atomic and Molecular Physics,}\\
\normalsize{Wuhan Institute of Physics and Mathematics, Chinese Academy of Sciences, Wuhan 430071, China}\\ 
\normalsize{$^{4}$State Key Laboratory of Surface Physics, Institute of Nanoelectronics and Quantum Computing,}\\
\normalsize{and Department of Physics, Fudan University, Shanghai 200433, China}\\
\normalsize{$^{5}$INO-CNR Istituto Nazionale di Ottica del CNR,}\\
\normalsize{Sezione di Sesto Fiorentino, I-50019 Sesto Fiorentino, Italy}\\
\normalsize{$^\star$E-mail: xuzongchen@pku.edu.cn, $^\ast$E-mail: xiaopeng\_li@fudan.edu.cn} 
} 

\date{\today}

\begin{document} 


\baselineskip24pt


\maketitle

\begin{sciabstract}
Turbulence is an intriguing non-equilibrium state, which originates from fluid mechanics and has far-reaching consequences in the description of climate physics, the characterization of quantum hydrodynamics, and the understanding of cosmic evolution. The concept of turbulent cascade describing the energy redistribution across different length scales offers one profound route to reconcile fundamental conservative forces with observational energy non-conservation of accelerating expansion of the universe bypassing the cosmological constant. Here, we observe a dimension crossing turbulent energy cascade in an atomic Bose-Einstein condensate confined in a two-dimensional (2d) optical lattice forming a 2d array of tubes, which exhibits universal behaviors in the dynamical energy-redistribution across different dimensions. By exciting atoms into the optical-lattice high bands, the excessive energy of this quantum many-body system is found to cascade from the transverse two-dimensional lattice directions to the continuous dimension, giving rise to a one-dimensional turbulent energy cascade, which is in general challenging to reach  due to integrability. We expect this observed novel phenomenon of dimension-crossing energy cascade may inspire microscopic theories for modeling positive cosmological constant of our inflationary universe. 
\end{sciabstract}

Since Leonardo da Vinci observed the structures of turbulence in fluid dynamics~\cite{2019_Kemp_Nature}, the study of this complex phenomenon has been attracting continuous attention in a broad context of mathematics and physics. Description and identification of universal properties of turbulence is an outstanding fundamental question owing to its intrinsic complexity to deal with macroscopic degrees of freedom across a wide range of length scales. Universal stationary cascade spectra of the energy distribution over different scales have been proposed for different types of turbulence. In the study of incompressible fluid, a universal Kolmogorov-Obukhov scaling form has been hypothesized for strong turbulence based on dimensional analysis~\cite{1941_Kolmogorov,1995_Frisch_turbulence}. For weak wave turbulence, a perturbative treatment with Zakharov conformal transformations has lead to a spectral form depending on the system dimensionality~\cite{zakharov1992statistical}.

The ubiquity of turbulence phenomena has been revealed in a variety of many-body systems, from classical liquids~\cite{lesieur1987turbulence,zakharov1992statistical,2017cardesaturbulent} to quantum gases~\cite{2016_Hadzibabic_Nature,Navon382,2009_Bagnato_PRL}, to astronomical scaled objects such as Great Red Spots on Jupiter~\cite{lesieur1987turbulence}. 
Recently the concept of turbulent energy cascade has also been applied to cosmology theories of our universe having an unexpected accelerating expansion, which gives rise to  the emergence of an effective cosmological constant that is the tendency of energy leakage from particles to the underlying Planck-scale spacetime~\cite{2013_Schillo_PLB,2017_Prez_PRL,2018_Perez_IJMP}.

\begin{figure}
	\center
	\includegraphics[width=\columnwidth]{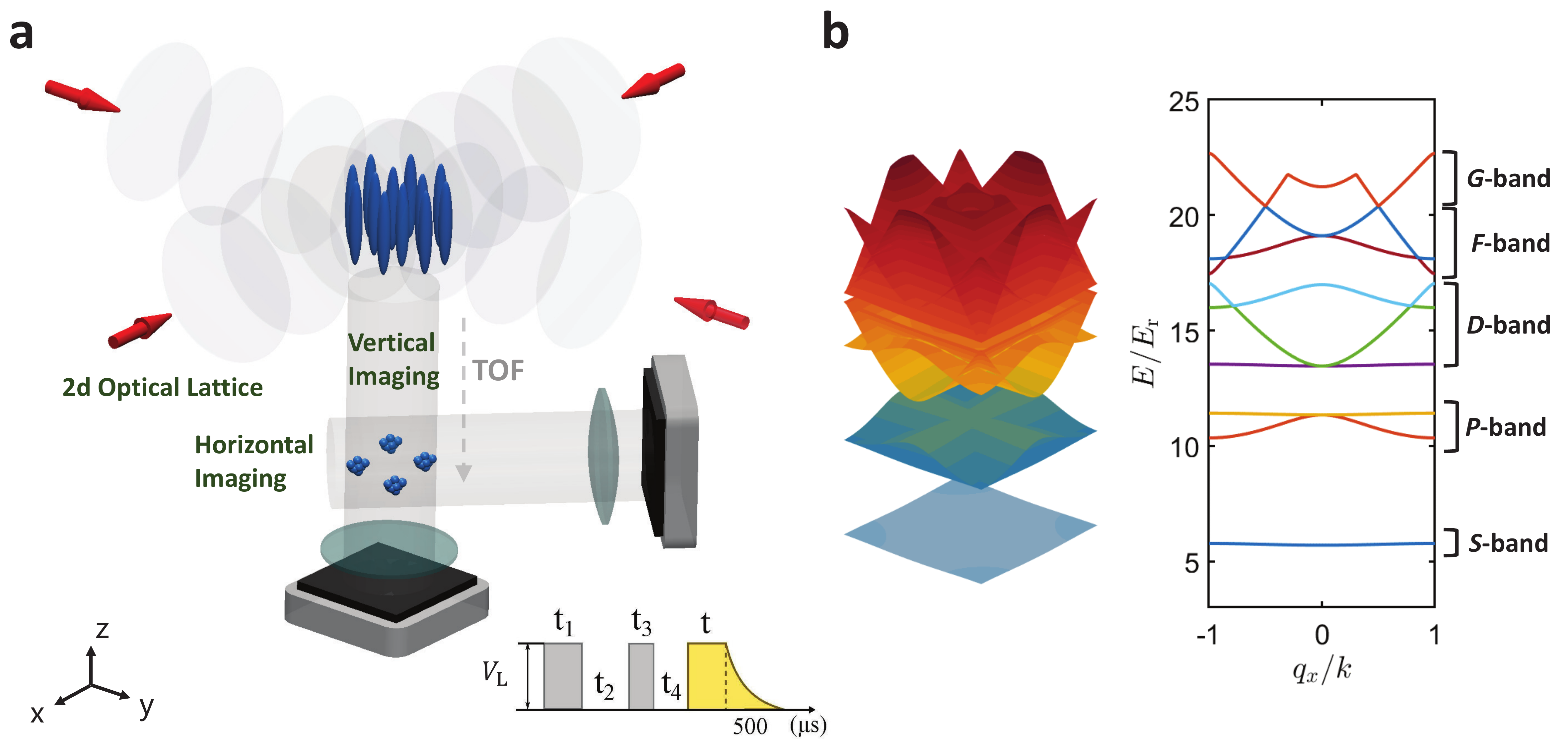}
	\caption{\textbf{Experimental setup to load atoms to excited bands of a tube lattice.} \textbf{a}, The lattice potential (red arrows) is created by two orthogonal retro-reflected laser beams in the horizontal plane. The BEC is loaded into the $G$-band of the 2d optical lattice through a shortcut loading procedure. After a variable lattice holding time $t$, atoms are released from the trap and imaged after a time-of-flight (TOF) either in the transverse directions (vertical imaging, to measure the band population) or in the longitudinal direction (horizontal imaging, to measure the momentum distribution). \textbf{b}, Schematics of Bloch bands of the 2d square lattice for the potential depth of $10\ E_{\rm r}$. The left panel shows the band structure from $S$- to $G$-band. The corresponding band energy along $x$ axis at $q_y=0$ is shown in the right panel.}
	\label{fig:fig1}
\end{figure}

Here we report an experimental observation of a stationary dimension crossing energy cascade in a 3d Bose-Einstein condensate (BEC) of $^{87}$Rb atoms confined in a 2d optical lattice (see Fig.~\ref{fig:fig1}). The atomic BEC is first prepared as the ground state of the 2d lattice, and then loaded into the high-orbital band acquiring excited band-energy in the transverse lattice directions (see Supplementary Information). The atoms then gradually decay back into the lowest band, where the excessive energy in the transverse directions smoothly leaks to the continuous dimension. During this dimension crossing energy cascade process, the momentum distribution along the longitudinal direction exhibits the scaling feature of one-dimensional energy-cascade of weak-wave turbulence, which has not been achieved in previous experiments due to the generic challenge caused by integrability of the one-dimensional weak-wave kinetic equation~\cite{zakharov1992statistical}. Considering the energy non-conservation of the subsystem in each optical tube analogous to our expanding universe, we expect the observed dimension crossing energy cascade may shed light on the construction of novel microscopic mechanism for positive cosmological constant~\cite{Riess_1998,1999_Permultter}.

The experiment starts with a BEC of $10^5$ $^{87}$Rb atoms in a nearly harmonic crossed-dipole trap with trapping frequency $2\pi\times\SI{20}{Hz}$~\cite{PhysRevA.99.013602}. The condensates are prepared in the $\lvert F=1,\,m_F=-1\rangle$ hyperfine state. A 2d optical lattice with a lattice spacing of $\SI{532}{nm}$ is formed by two orthogonal retro-reflected laser beams in the horizontal plane (see Fig.~\ref{fig:fig1}a). The atoms are confined in about 1500 tubes by the lattice~\cite{2018arXiv180401969L,Haller1224,li2016physics,2015_Dutta_Lattice}. We develop a fast band-loading technique for the 2d optical lattice to prepare excited-band BEC~\cite{li2016physics,Kock_2016} by generalizing a 1d shortcut method~\cite{LiuXX, ZhaiYY, HuDong, PhysRevLett.121.265301,1367-2630-20-5-055005}. 
Using this technique, the system is excited to the zero quasi-momentum state of the $G$-band after a ground state BEC is prepared. After a variable holding time of $t$, the lattice is exponentially ramped down within $\SI{500}{\mu s}$, we measure both of the band population along the lattice directions (denoted as the $xy$-directions) and the momentum distribution in the continuous direction (the $z$-direction). The atomic distribution in different Brillouin zones is probed by the imaging laser beam propagating along the $z$ axis (see vertical imaging in Fig.~\ref{fig:fig1}a) following a band mapping procedure~\cite{RN2341}. The momentum distribution in the continuous direction is probed by the imaging laser beam propagating along the angular bisector between the $x$- and $y$-axis (horizontal imaging in Fig.~\ref{fig:fig1}a).

\begin{figure}
	\center
	\includegraphics[width=\columnwidth]{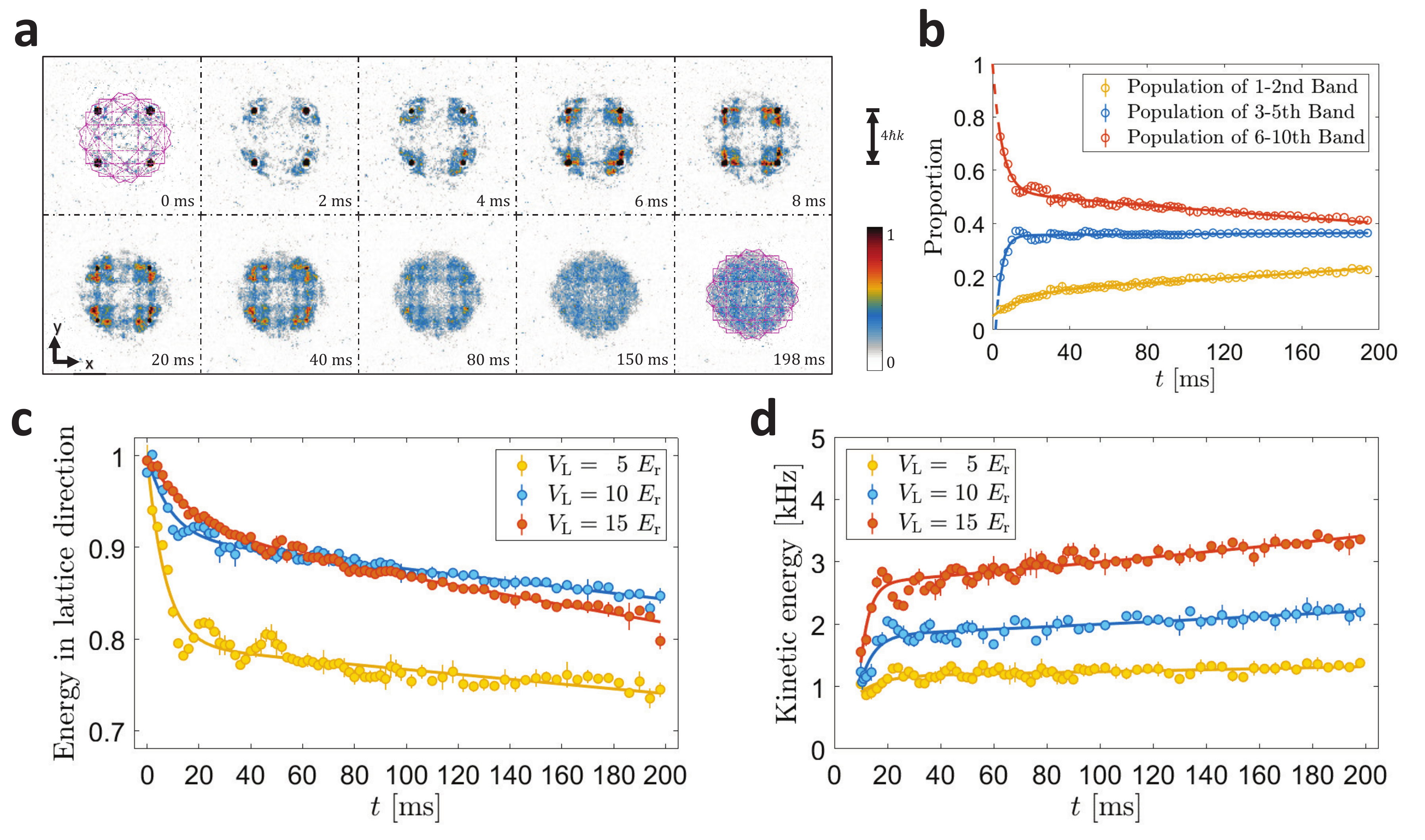}
	\caption{\textbf{Time evolution of the band relaxation dynamics and the dimension crossing energy transfer.} \textbf{a}, Mapping of band populations with increasing lattice holding times for the potential depth of $10\ E_{\rm r}$. The pink lines illustrate the first nine Brillouin zones of the 2d square lattice (Supplementary Information). \textbf{b}, The time evolution of different band populations with the lattice depth of $10\ E_{\rm r}$. \textbf{c}, The time evolution of the energy along the transverse directions with different lattice depths. The energy is rescaled by the initial energy of each lattice depth at $t=\SI{0}{ms}$. The solid lines show fitting of the experimental  data using a double exponential decay form that has two time scales $\tau_1$ and $\tau_2$ (Eq.~\eqref{eq:fitting}). For the lattice depths of $V_{\rm L}/E_{\rm r} = 5,\,10,\,15$, we have $[\tau_1=\SI{6.95(4)}{ms}$, $\tau_2=\SI{286(0)}{ms}]$, $[\tau_1=\SI{8.96(1)}{ms}$, $\tau_2=\SI{258(6)}{ms}]$, and $[\tau_1=\SI{17.2(4)}{ms}$, $\tau_2=\SI{167(5)}{ms}]$, respectively. The error bars denote the standard deviation for averaging five experimental images. \textbf{d}, The time evolution of the kinetic energy along the longitudinal direction with different lattice depths.}
	\label{fig:fig2}
\end{figure}

After the BEC is loaded onto the $G$-band of the lattice, it becomes dynamically unstable, and starts to decay into the lower bands (see Fig.~\ref{fig:fig2}). In the deep lattice limit, the decay rate to the ground band (per unit length in each tube) is given by Fermi's Golden rule at leading order (Supplementary Information), 
\be 
w_t \propto \frac{\hbar^3 }{[M a_{\rm 1d} ^{(SG)} l_{\rm 1d}]^2}  \rho (\Delta E),  
\ee
with $a_{\rm 1d} ^{(SG)}$ the 1d scattering length~\cite{1998_Olshanii_PRL} describing the interaction process between atoms populating $S$- and $G$-orbitals in the tube, $l_{\rm 1d}$ the average inter-particle distance, $\Delta E$ the energy release from the $G$- to the $S$- band, and the density of states $\rho(\Delta E) = \int \frac{dk_z}{2\pi} \delta (\hbar^2  k_z^2 /2M - \Delta E)$. 
Having continuous degrees of freedom in the longitudinal direction allows atoms to decay into a final state with arbitrary transverse lattice momenta, which is observed in the experiment (Fig.~\ref{fig:fig2}a). 
In Fig.~\ref{fig:fig2}a-c, the time evolution of the band relaxation dynamics is shown. Once loaded into the $G$-band, atoms mainly reside on the four transverse momenta ($\pm2\hbar k$, $\pm2\hbar k$), which correspond to the zero lattice-momentum state. 
We find the decay of the $G$-band population fits well to an empirical form of (see Fig.~\ref{fig:fig2}c), 
\be 
N_G (t) = A \exp (-t/\tau_1) + B\exp(-t/\tau_2), 
\label{eq:fitting} 
\ee
which implies two time scales ($\tau_1 \ll \tau_2$) or two evolution stages in the band relaxation dynamics.  
At the first stage within a few to twenty milliseconds, atoms exhibit a fast decay to lower bands, while the sharp peaks at the four momenta ($\pm2\hbar k$, $\pm2\hbar k$) remain visible indicating the maintenance of the phase coherence across different tubes. The energy loss in the resultant band relaxation is compensated by the energy absorbed by the continuous degrees of freedom in the tube.  We find the characteristic time scale of the decay dynamics ($\tau_1$) increases as we increase the lattice depth, which is qualitatively consistent with the decrease in the 1d density of states as the energy release in the band relaxation $\Delta E$ increases at a larger lattice depth.  
At the second stage, the phase coherence disappears and the decay dynamics becomes slower. After about $\SI{200}{ms}$, the atomic population among the different Brillouin zones becomes almost uniform, which means the phase coherence across the tubes fades away almost completely.

\begin{figure}
	\center
	\includegraphics[width=\columnwidth]{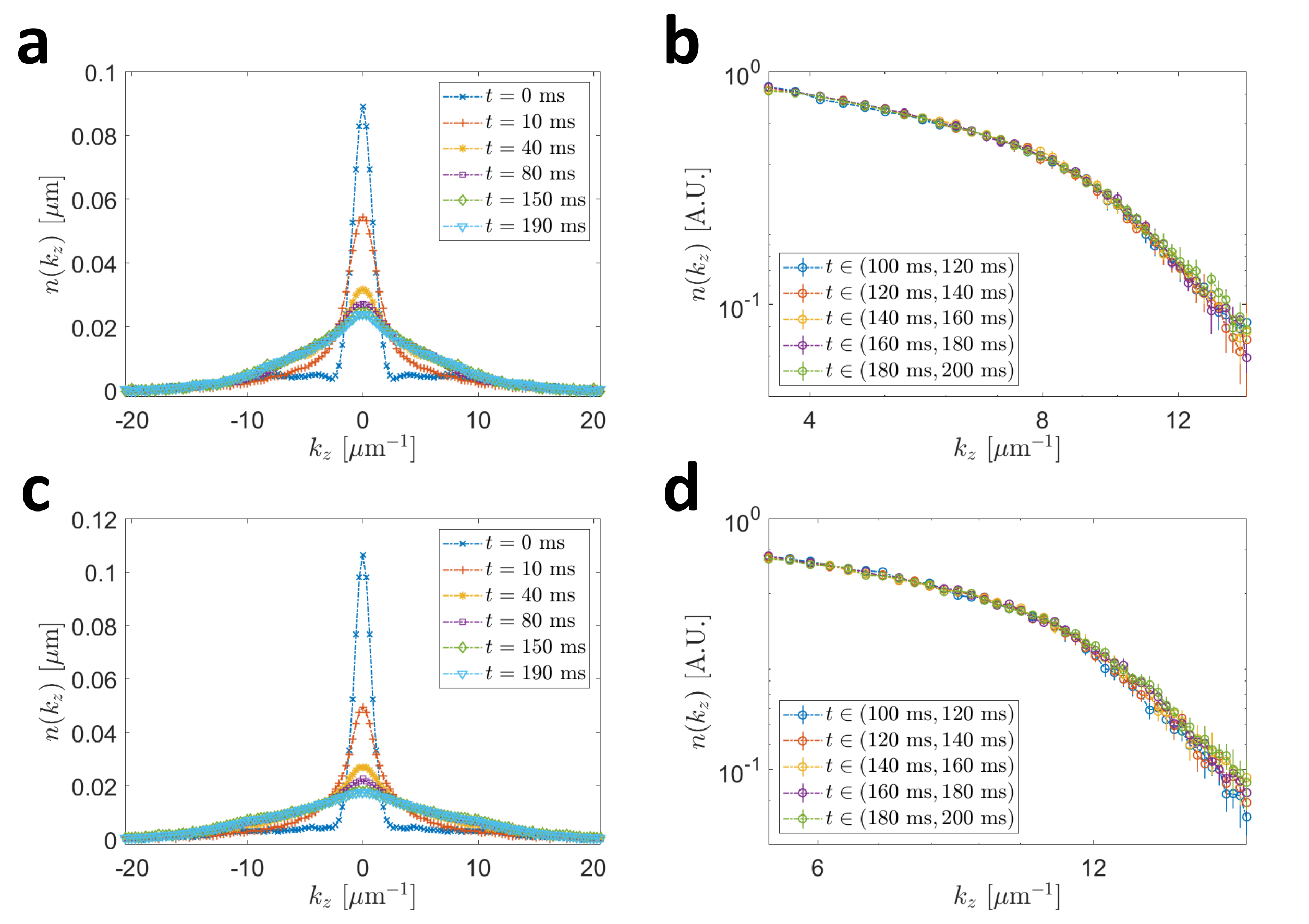}
	\caption{\textbf{Dynamical evolution of the momentum distribution along the continuous dimension.} 
	{\bf a},{\bf c}, The normalized momentum distribution at different lattice holding times. The momentum distribution $n(k_z)$ gradually broadens out from the form of a condensate and approaches a universal form in the dynamical evolution. {\bf b},{\bf d}, The averaged momentum distributions (in log-log scale) at different time windows as shown in the figure. In {\bf b}, {\bf d}, the momentum distribution is normalized by the atom number in the range of $k_z \in (\SI{2}{\mu m^{-1}} , \SI{21}{\mu m^{-1}})$ for illustration purpose. We have lattice depths---$10\ E_{\rm r}$ and $15\ E_{\rm r}$ in {\bf a},{\bf b}, and {\bf c},{\bf d}, respectively.}
	\label{fig:fig3}
\end{figure}

During the cross-dimensional energy cascade from the lattice to the continuous dimension, we find the momentum distribution along the $z$ direction $n(k_z)$ exhibits universal behaviors (Fig.~\ref{fig:fig3}). Right after the transverse high band is populated, the momentum distribution $n(k_z)$ still resembles the form of a BEC, which means the superfluid phase coherence preserves. By increasing the holding time, we observe $n(k_z)$ starts to broaden out in dynamical evolution, deviating from the form of a condensate. This can be attributed to band-excitation energy converting to the kinetic energy along the tube. 
Following the time evolution, we find the momentum distribution $n(k_z)$ approaches a universal form. This behavior is found to emerge for both lattice depths of $V_{\rm L}=10\ E_{\rm r}$ and $V_{\rm L}=15\ E_{\rm r}$, which implies it is a generic phenomenon.

\begin{figure}
	\center
	\includegraphics[width=\columnwidth]{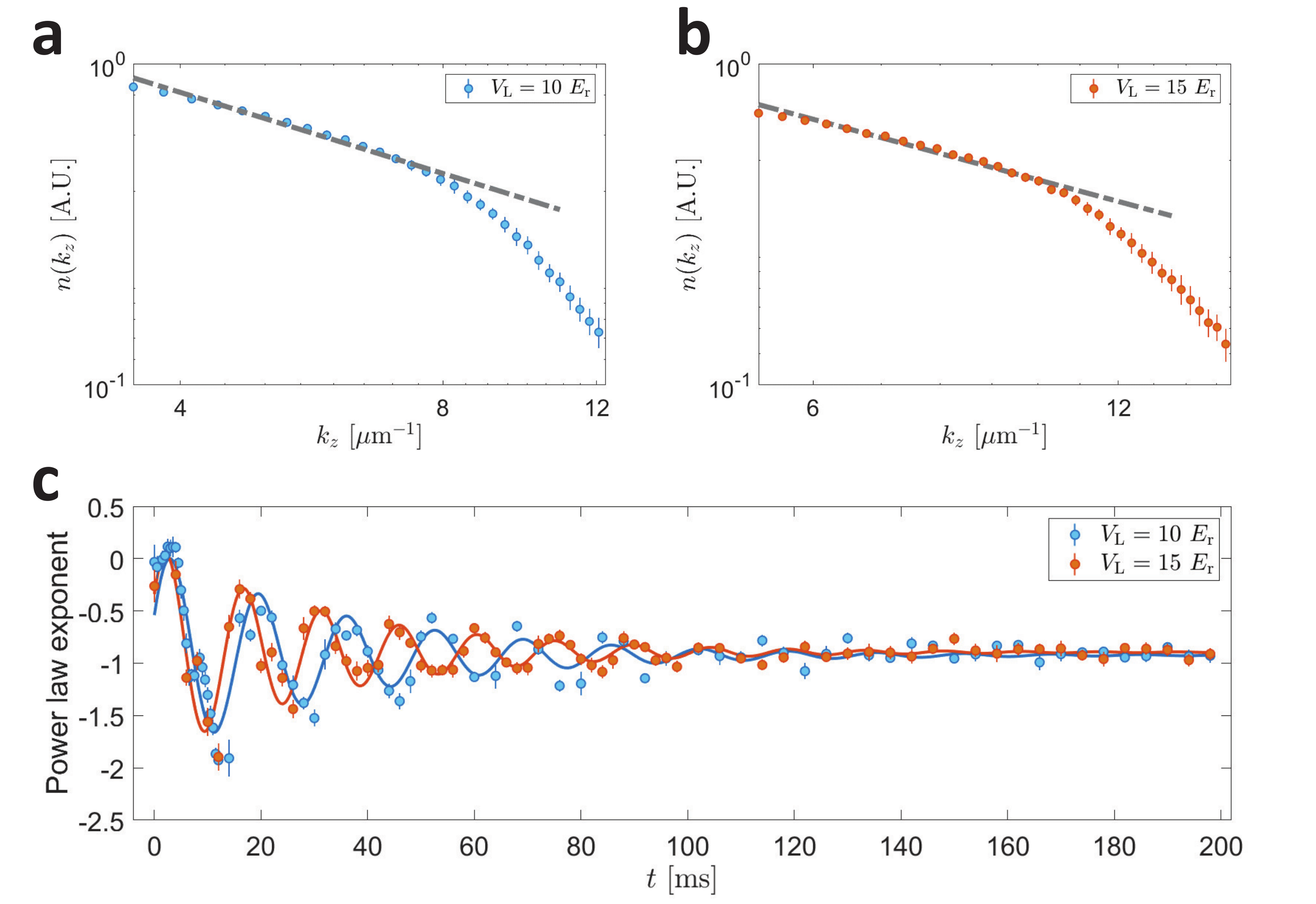}
	\caption{\textbf{Universal turbulent cascade spectrum.} {\bf a},{\bf b}, The late-time (after $\SI{100}{ms}$) averaged momentum distribution in log-log scale. The error bars denote the standard deviation of the data points. The `grey dashed' lines represent power law fit of the momentum distribution. The power-exponent is $-0.84\pm0.01$ for the lattice depths of $10\ E_{\rm r}$, and $-0.85\pm0.01$ for the lattice depths of $15\ E_{\rm r}$ with the error obtained through the statistical Bootstrap method. \textbf{c}, The dynamical evolution of the power law exponent for the lattice depths of $10\ E_{\rm r}$ and $15\ E_{\rm r}$.}
	\label{fig:fig4}
\end{figure}

We then average over momentum distributions at late times, and the results are shown in Fig.~\ref{fig:fig4}a and Fig.~\ref{fig:fig4}b.  
The averaged momentum distribution exhibits a wide range  of power law distribution with a power-exponent about $-0.85$. This exponent essentially remains the same for different choices of lattice depths. Fig.~\ref{fig:fig4} shows the experimental results for $V_{\rm L}=10\ E_{\rm r}$ and $V_{\rm L}=15\ E_{\rm r}$. This universal power-law decay observed in our non-equilibrium setting represents the emergence of a one-dimensional energy cascade spectrum in a three-dimensional dynamical system. We emphasize that this observed one-dimensional energy spectrum cannot emerge in a truly one-dimensional system, due to the integrability of the weak-wave turbulence equation in one-dimension. Our experiment thus reveals a novel way bypassing this integrability in achieving the universal one-dimensional energy cascade spectrum. The experimental observations are also qualitatively distinctive from the quench dynamics of an isolated 1d Bose gas~\cite{2018_Schmiedmayer_universal}. We note that the observed turbulence spectrum exponent has a slight deviation from dimension analysis, which is further confirmed in our numerical simulation (Supplementary Information). This level of discrepancy has also been found in previous studies of three-dimensional weak-wave turbulence~\cite{2016_Hadzibabic_Nature,Navon382,2016_Matsumoto_PRE}. 

To characterize the dynamical evolution, we also analyze the momentum distributions at individual evolution times. The dynamical evolution of the resultant exponents are shown in Fig.~\ref{fig:fig4}c. The system does not form stationary energy cascade immediately after we load the BEC to the $G$-band, but instead first undergoes a while of non-universal oscillations before the universal stationary cascade emerges.

It is worth remarking that the power-law decay in the experiment cannot be described by the equilibrium theory of quantum Luttinger liquid~\cite{giamarchi2003quantum,2004_Bloch_Nature}, because the observed power-exponent is universal whereas it should vary for different lattice depths due to interaction effects in Luttinger liquids~\cite{giamarchi2003quantum,2004_Bloch_Nature,RN2342}.

\paragraph*{Discussion and Outlook.} 
Our experiment discovers a novel type of turbulent cascade for the dynamical energy redistribution across different dimensions. In this dimension crossing energy cascade, we have observed a universal one-dimensional turbulence behavior in the momentum distribution along the continuous dimension, which is difficult to achieve in other setups due to integrability of the one-dimensional weak wave turbulence kinetic equation. In addition, the observed phenomenon of dimension crossing energy cascade may inspire novel microscopic mechanism for cosmic evolution with positive cosmological constant.

\paragraph*{Acknowledgement.} 
We acknowledge helpful discussion with Maciej Lewenstein. 
This work is supported by National Program on Key Basic Research Project of China (Grant No. 2016YFA0301501, Grant No. 2017YFA0304204), 
National Natural Science Foundation of China (Grants No. 91736208, No. 11920101004, No. 61727819, No. 61475007, No. 11934002, No. 11774067, No. 61703025, and No. 11504328). 

\bibliographystyle{naturemag}
\bibliography{Reference}

\paragraph*{Author Contributions} 
All the authors contributed to the elaboration of the project and helped in editing the manuscript.

\paragraph*{Competing Interests} 
The authors declare that they have no competing financial interests.


\newpage 

\renewcommand{\theequation}{S\arabic{equation}}
\renewcommand{\thesection}{S-\arabic{section}}
\renewcommand{\thefigure}{S\arabic{figure}}
\renewcommand{\thetable}{S\arabic{table}}
\setcounter{equation}{0}
\setcounter{figure}{0}
\setcounter{table}{0}

\clearpage 

\begin{center}
{\bf \Huge Supplementary Materials}
\end{center}

\section{Shortcut method to load atoms into $G$-band}

The shortcut loading method is an efficient procedure for transferring a BEC from the ground state of a harmonic trap into the designated bands of an optical lattice. It is based on a designed lattice laser pulse sequence completed within several tens of microseconds. We consider the general situation of a 1d optical lattice, when neglecting the atom-atom interactions (because the loading time is very short), the single-atom Hamiltonian is then given by
\be
\hat{H}_{\rm L}=-\frac{\hbar^2\nabla^2}{2M}+V_x\cos^2(k_x x),
\ee
where $V_x$ is the lattice depth and $k_x$ the lattice laser wave number. The eigenstates of the Hamiltonian $\hat{H}_{\rm L}$ can be expressed as the Bloch states ($\hbar=1$)
\be
\lvert n, q\rangle=u^{(n)}_q\mathrm{e}^{-\mathrm{i}qx}=\sum_{\ell}c_{n,\ell}\lvert 2\ell k_x+q\rangle,
\ee
with the band index $n$, the quasi-momentum $q$ and $\ell=0,\,\pm 1,\,\pm 2 \cdots $. For transferring a BEC with initial wave-function $\lvert\psi_{\rm i}\rangle$ into the $D$-band (second excited band) with zero quasi-momentum, the target state $\lvert\psi_{\rm t}\rangle$ is $\lvert n=3, q=0\rangle$. We apply a four-step lattice laser pulse sequence before switching on the optical lattice with the potential depth of $V_x$, the state after applying the lattice laser pulse sequence is
\be
\lvert\psi_{\rm f}\rangle=\prod^{4}_{j=1}\hat{U}_j\lvert\psi_{\rm i}\rangle,
\ee
where $\hat{U}_j=\mathrm{e}^{-\mathrm{i}\hat{H}_j t_j/\hbar}$ is the evolution operator in the $j$th step, $\hat{H}_j$ is the Hamiltonian corresponding to the lattice depth $V_j$ and $t_j$ is the pulse duration. We fix the lattice depth to $V_j=V_x$ or 0, so in the pulse sequence $[t_1,\,t_2,\,t_3,\,t_4]$ corresponds to the lattice depth $[V_x,\,0,\,V_x,\,0]$. The parameters $\hat{H}_j$ and $t_j$ can be determined by maximizing the fidelity
\be
F=\lvert\langle\psi_{\rm t}\lvert\psi_{\rm f}\rangle\lvert^2.
\ee
By properly choosing the values of $[t_1,\,t_2,\,t_3,\,t_4]$, the BEC can be transferred into the $D$-band of the 1d optical lattice for different potential depths with close to $100\%$ fidelity.

\begin{figure}
	\center
	\includegraphics[width=\columnwidth]{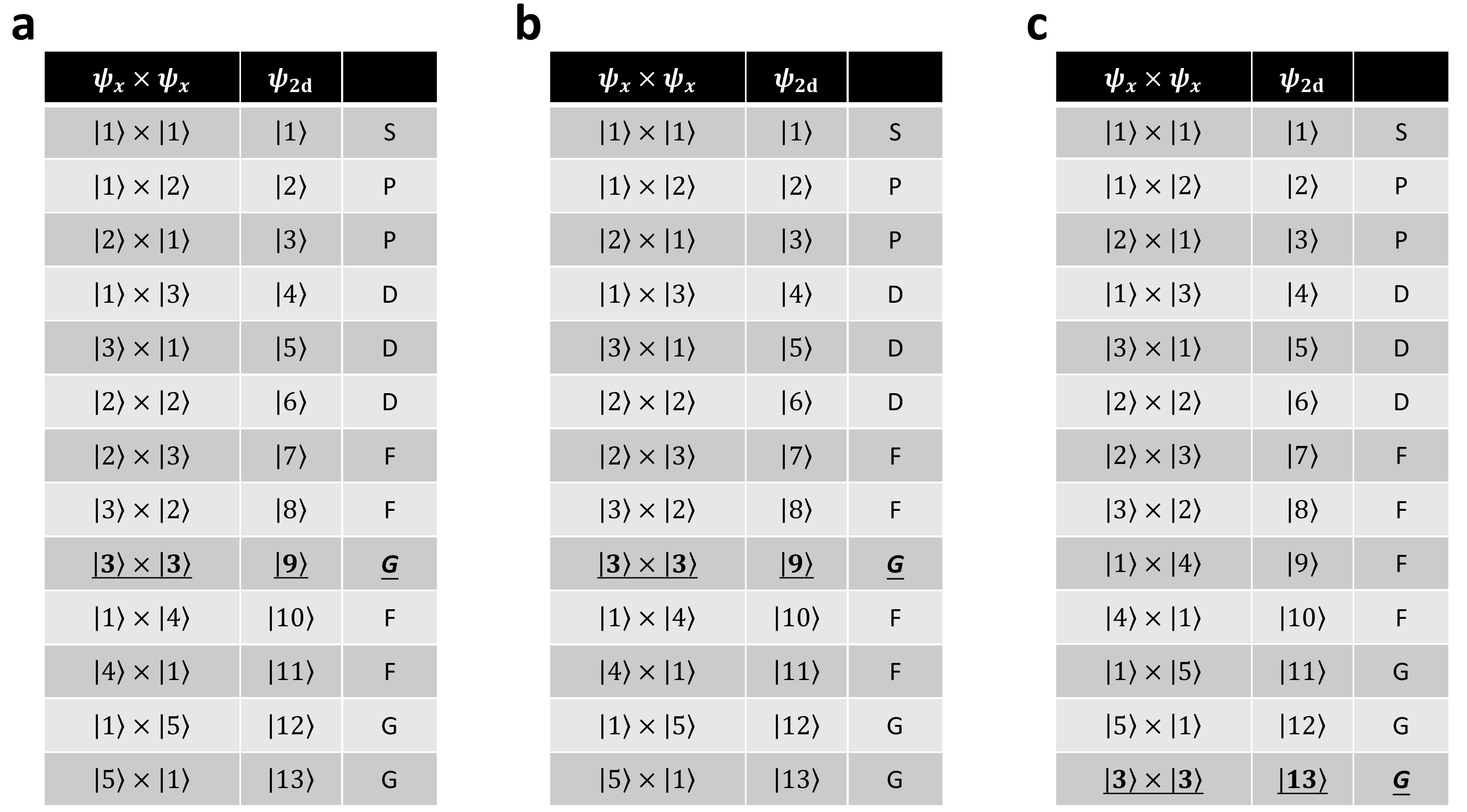}
	\caption{The corresponding relations between the products of two 1d Bloch states and the Bloch states of 2d square lattices with zero quasi-momentum for different lattice depths. Values of the lattice depths: {\bf a}, $V_x=V_y=5\ E_{\rm r}$; {\bf b}, $V_x=V_y=10\ E_{\rm r}$; and {\bf c}, $V_x=V_y=15\ E_{\rm r}$. In actual calculation, there is a slight difference of $0.01\ E_{\rm r}$ between $V_x$ and $V_y$, in order to avoid energy degeneracy at zero quasi-momentum.}
	\label{fig:figS1}
\end{figure}

This procedure can be extended to a 2d coordinate separable square lattice. In this case, the wave functions can be separated in the form of $\psi_{\rm 2d}=\psi_x\psi_y$ and the evolution operator can be separated as $\hat{U}_{\rm 2d}\psi_{\rm 2d}=\hat{U}_x\psi_x\hat{U}_y\psi_y$. The corresponding relations between the products of two 1d Bloch states and the Bloch states of 2d square lattices for potential depths $5\ E_{\rm r}$, $10\ E_{\rm r}$ and $15\ E_{\rm r}$ are shown in Fig.~\ref{fig:figS1}.

\section{Measurement of momentum distributions} 

We use standard TOF techniques to detect the momentum distribution along the longitudinal direction, by resonant absorption imaging after $t_0=\SI{30}{ms}$ of free expansion. We then normalize the integrated column density $\hat{n}(k_z)$ measured after lattice holding time $t=\SI{100}{ms}$ by the atom number in the range of $k_z \in (\SI{2}{\mu m^{-1}} , \SI{21}{\mu m^{-1}})$, since when the universal momentum distribution has formed. From the sampled datasets consisting of the momentum distributions at different lattice holding times, we randomly resample the momentum distribution profiles without avoiding possible repeating, and perform power law fitting on their average. We do this resampling-fitting for $10^3$ times, then evaluate the power-exponent by taking the mean value and the standard deviation according to the statistical Bootstrap method.

\section{Fermi's Golden rule transition rate} 

The band relaxation of the excited BEC corresponds to Fermi's Golden rule transition. The interaction of the system reads 
\be 
H_{\rm int} = \frac{2\pi \hbar^2 a_{\rm s}}{M} \int d^3 {\bf x} \phi ^\dag ({\bf x}) \phi^\dag ({\bf x}) \phi({\bf x}) \phi({\bf x}), 
\label{seq:Hint} 
\ee 
with $\phi({\bf x})$ the bosonic field operator.  This operator can  be expanded by the degrees of freedom in the tube as 
\be 
\phi({\bf x}) = \frac{1}{\sqrt{N_t L_z}} \sum_{\alpha,{\bf k}} b_\alpha ({\bf k}) \varphi_{\alpha,{\bf k}_\perp} ({\bf x}_\perp) \mathrm{e}^{\mathrm{i}k_z z + \mathrm{i}{\bf k}_\perp \cdot {\bf x}_\perp}, 
\ee 
with $N_t$ the number of tubes, and $L_z$ the length of each tube, $b_\alpha ({\bf k})$ the annihilation operator with momentum ${\bf k} = (k_x, k_y, k_z)$ and the band index $\alpha$, and $\varphi_{\alpha,{\bf k}_\perp} ({\bf x}_\perp)$ the lattice-periodic Bloch function. Here  we have also introduced a shorthand notation ${\bf x}_\perp = (x,y)$ for the spatial coordinates in the transverse direction , and  ${\bf k}_\perp = {k_x, k_y}$ for the lattice momentum. We have neglected harmonic trap potential in this estimate of band relaxation rate. The multi-band interaction processes can be derived by rewritting the interaction in Eq.~\eqref{seq:Hint} in terms of $b_\alpha ({\bf k})$ operators. 
With the BEC excited to the $G$-band described as 
$|\Psi\rangle =\frac{[b_G ^\dag (0) ]^N}{\sqrt{N!}} |{\rm vac} \rangle$, the decay rate (per unit length in each tube) to the lowest band according to the Fermi's Golden rule is given as 
\be 
w_t \propto \frac{\hbar^3 }{[M a_{\rm 1d} ^{(SG)} l_{\rm 1d}]^2}   [N_t L_z ]^{-1} \sum_{\bf k} \delta \left( \epsilon_G (0) - \epsilon_S ({\bf k}_\perp) - \frac{\hbar^2 k_z ^2 }{2M} \right), 
\ee 
with $l_{\rm 1d}$ the average inter-particle distance, $a_{\rm 1d} ^{(SG)}$ the 1d scattering length~\cite{1998_Olshanii_PRL} describing the interaction process between the $S$- and $G$-band particles in the tube. In the deep lattice limit, the band dispersion becomes negligible compared to the band gap. Then the transition rate can be written in terms of the 1d density of states (per unit length) as 
\be 
w_t \propto \frac{\hbar^3 }{[M a_{\rm 1d} ^{(SG)} l_{\rm 1d}]^2}  \rho (\Delta E), 
\ee 
with $\Delta E$ the energy release for the decay from the $G$- to $S$-band, and the density of states $\rho(\Delta E) = \int \frac{dk_z}{(2\pi)} \delta \left( \Delta E-\frac{\hbar^2k_z ^2}{2M} \right)$.  
This analysis also leads to a characteristic momentum $k_{zF} = \sqrt{2M\Delta E /\hbar^2 } $ at which atoms would immediately populate after decay from the higher $G$-band. 

\begin{figure}[htp]
	\center
	\includegraphics[width=\columnwidth]{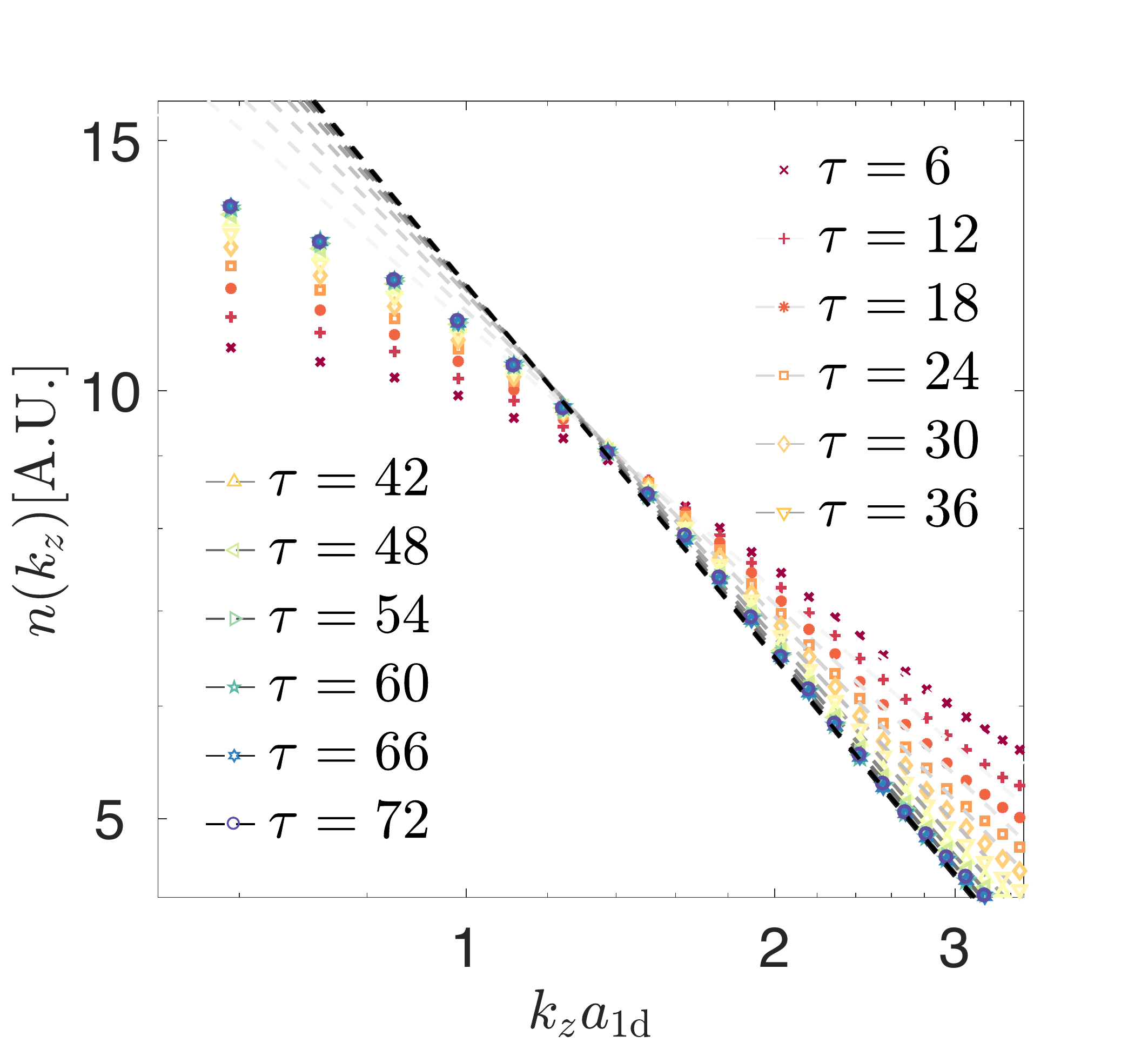}
	\caption{Numerical results for quasi one dimensional energy cascade. The momentum distribution shown here is normalized. We simulate bosons in a 2d array of $16\times16$ tubes with time-dependent Gross-Pitaevskii equation. The 1d scattering length is set to be $|a_{\rm 1d}|=\SI{200}{nm}$. The inter-tube tunneling is set to be $t_\perp = E_{\rm r}/10$, with the single-photon recoil energy $E_{\rm r} = \frac{2 \pi^2 \hbar^2}{M\lambda^2}$, the wavelength $\lambda = \SI{1064}{nm}$. The non-hermitian term $\mu_I  = 0.5 t_\perp$. The time $\tau$ in the plot is dimensionless as scaled by $\hbar/t_\perp$.  The late time dynamics corresponds to a one-dimensional energy cascade with a power-law decaying momentum distribution, $n(k_z) \sim k_z ^\eta$. We find $\eta \approx -0.9$ in our numerical simulation, which agrees with the 1d turbulent cascade but at the same time has a slight deviation from the dimensional analysis.}
	\label{fig:fig5}
\end{figure}

\section{Emergence of 1d turbulent energy cascade in the tube-lattice}
 
The scaling analysis of weak-wave turbulence predicts a power-law decaying momentum distribution $1/k_z$ in one-dimension. However, this turbulent energy cascade cannot be reached in a truly 1d system due to the integrability of the weak-wave kinetic equation. In this section, we construct an effective description and show that the 1d energy cascade could emerge in the tube-lattice. Consider the dynamics of annihilation operator of the lowest band in the Heisenberg picture, $b_{\bf r} (k_z)$, with ${\bf r}$ and $k_z$ the coordinate of each tube and the longitudinal momentum. We have 
\bea 
i\partial_t b_{\bf r} (k_z) &=& -t_\perp \sum_{{\bf r}'\in \partial {\bf r}} b_{{\bf r}'} (k_z)  + \frac{\hbar^2 k_z ^2}{2M} b_{\bf r}(k_z)  
+ i\mu_I  b_{\bf r} (k_z) \left [ \delta_{k_z, k_{zF}} +  \delta_{k_z, -k_{zF}} \right] \nn \\
&+& 2 g_{\rm 1d}/L_z \sum_{\bf r} \sum_{q_z, q_z' }   b_{\bf r} ^\dag ( q_{z} + q_{z}'-k_z ) b_{\bf r} (q_z) b_{\bf r} (q_z '). 
\eea
Here $t_\perp$ is the inter-tube tunneling, $\mu_I$ models the particles scattered into the lowest band corresponding to  the Fermi's Golden rule picture, and $g_{\rm 1d} = -\frac{2\hbar^2}{M a_{\rm 1d}}$ ($a_{\rm 1d}$ the 1d scattering length) describes the interaction effects in each tube~\cite{1998_Olshanii_PRL}. 

Taking the Gross-Pitaevskii approach, we replace the annihilation/creation operators with their mean field values, which is valid at weak interaction. For the initial state, we choose a condensate wavefunction with a random phase because the scattered particles from the excited to the lowest band are not expected to be phase-coherent. In our numerical calculation, we average over $10^3$ different initial phase configurations. The results are shown in Fig.~\ref{fig:fig5}. We find that the shape of the momentum distribution along the continuous dimension asymptotically approaches to a universal function $k_z ^{-0.9}$ at late time in absence of inter-tube phase coherence. This behavior is close to the 1d turbulent energy cascade from the dimensional analysis with a slight difference. We note here that the discrepancy between numerics and dimensional analysis has also been found in previous literatures~\cite{2016_Hadzibabic_Nature,Navon382,2016_Matsumoto_PRE}. 

\end{document}